\documentclass{aa}

\usepackage{graphicx}
\usepackage{txfonts}
\usepackage{xcolor}
\usepackage{amsmath}
\usepackage[breaklinks=true, colorlinks=true, citecolor=blue, linkcolor=blue,urlcolor=blue]{hyperref}

\usepackage{siunitx}
\DeclareSIUnit\parsec{pc}
\DeclareSIUnit\gauss{G}
\DeclareSIUnit\erg{erg}
\DeclareSIUnit\year{yr}
\DeclareSIUnit\arcmin{arcmin}
\DeclareSIUnit\arcsec{arcsec}
\DeclareSIUnit\counts{cts}
\DeclareSIUnit\jansky{Jy}
\DeclareSIUnit\steradian{sr}
\DeclareSIUnit\dex{dex}

\usepackage{orcidlink}

\usepackage{ctable}

\newcommand{\hi}{{\rm H}\,{\small\rm I}}

\begin{document}

    \title{A new uncertainty scheme for galaxy distances from flow models}

    \author{
            Konstantin~Haubner\orcidlink{0009-0007-7808-4653}\inst{\ref{inst1},\ref{inst2}}
        \and
            Federico~Lelli\orcidlink{0000-0002-9024-9883}\inst{\ref{inst1}}
        \and
            Enrico Di~Teodoro\orcidlink{0000-0003-4019-0673}\inst{\ref{inst2},\ref{inst1}}
        \and
            Francis~Duey\orcidlink{0009-0003-1662-5179}\inst{\ref{inst3}}
        \and
            Stacy~McGaugh\orcidlink{0000-0002-9762-0980}\inst{\ref{inst3}}
        \and
            James~Schombert\orcidlink{0000-0003-2022-1911}\inst{\ref{inst4}}
}
    \institute{
            INAF -- Arcetri Astrophysical Observatory, Largo Enrico Fermi 5, 50125 Firenze, Italy\\
            \email{konstantin.haubner@inaf.it}\label{inst1}
        \and
            Dipartimento di Fisica e Astronomia, Università degli Studi di Firenze, via G. Sansone 1, 50019 Sesto Fiorentino, Firenze, Italy\label{inst2}
        \and
            Department of Astronomy, Case Western Reserve University, Cleveland, OH 44106, USA\label{inst3}
        \and
            Department of Physics, University of Oregon, Eugene, OR 97403, USA\label{inst4}
}

    \date{Received 17 February 2025 / Accepted 11 March 2025}

    \abstract{
      The systemic velocity or redshift of galaxies is a convenient tool to calculate their distances in the absence of primary methods, but the uncertainties on these flow distances may be substantial due to galaxy peculiar motions. Here, we derived a simple and easily applicable method to assign uncertainties to flow distances from four different methodologies, namely the Hubble law with both heliocentric and local-sheet velocities, the Cosmicflows-4 model, and the numerical action methods model. Our uncertainty scheme was constructed by comparing these flow distances to accurate, redshift-independent distances of a subsample of ${\sim}2000$ galaxies from the Cosmicflows-4 database, using the tip magnitude of the red giant branch, Cepheids, surface brightness fluctuations, supernovae type Ia, masers, and supernovae type II. We provide simple functions and tables to calculate the distance uncertainties for all the flow models considered. This uncertainty scheme is generally applicable except for the region around the Virgo cluster, where we assign increased uncertainties due to larger peculiar motions.
    }

    \keywords{methods: miscellaneous --
            galaxies: distances and redshifts
    }

    \maketitle

\nolinenumbers

\section{Introduction}

Distances are one of the most important quantities in astrophysics because they are necessary for the derivation of many intrinsic physical properties of astronomical objects. For determining distances, we typically resort to objects whose absolute luminosity is known or can be calculated through correlations with distance-independent observables. Commonly used examples are the tip magnitude of the red giant branch (TRGB; \citeauthor{Lee1993} \citeyear{Lee1993}), the Cepheid period-luminosity relation (CPLR; \citeauthor{Leavitt1912} \citeyear{Leavitt1912}), supernovae type Ia (SNIa) through their standardizable light curves \citep{Phillips1993}, and the plateau subclass of supernovae type II (SNII) through a correlation between their expansion velocity and luminosity \citep{Hamuy2002}.

Although primary distances\footnote{We use the term "primary distance" to refer to any distances determined with redshift-independent methodologies.} derived through such standard candles often have excellent accuracies of $5$ to $10\%$ \citep{Tully2009}, they are not available for the vast majority of extragalactic objects. For example, one of the largest assemblies of primary distances on a homogeneous distance scale, the Cosmicflows-4 (CF4) database, has entries for roughly $56000$ galaxies \citep{Tully2023}. Of these distances, around $54000$ were derived using kinematic scaling relations, namely (1) the classic Tully-Fisher relation (TFR; \citeauthor{Tully1977} \citeyear{Tully1977}), a correlation between the \hi\ line width (a proxy for rotation velocity) and the luminosity of disk galaxies, (2) the underlying baryonic Tully Fisher relation (BTFR; \citeauthor{McGaugh2000} \citeyear{McGaugh2000}; \citeauthor{Lelli2019} \citeyear{Lelli2019}), a correlation between the flat rotation velocity and the total baryonic mass of disk galaxies, or (3) the fundamental plane (FP; \citeauthor{Dressler1987} \citeyear{Dressler1987}; \citeauthor{Djorgovski1987} \citeyear{Djorgovski1987}), a correlation between the luminosity, surface brightness, and velocity dispersion of elliptical galaxies. However, these kinematic relations are often studied in their own right to constrain models of galaxy formation and evolution, the properties of galactic dark matter halos, and theories of modified dynamics \citep[e.g.,][]{Verheijen2001, Cappellari2013, Lelli2016b, Desmond2019}. If the goal is to investigate these scaling relations to understand the underlying physics, one should not use the same relations for the determination of the required galaxy luminosities and masses. Such a procedure would impose galaxies to follow these relations by construction, which would bias the results. Consequently, if one is interested in kinematic studies of individual galaxies, the number of primary distances available in CF4 drops to roughly $2000$.

In the absence of primary distances, it is common practice to use the Hubble law, which links the distance of a galaxy to its observed redshift \citep{Hubble1929}. However, this method suffers from the degeneracy between redshift due to the expansion of the Universe and redshift due to the peculiar motions of galaxies. Peculiar velocities can reach values of several $\SI{100}{\kilo\meter\per\second}$ \citep{Tully2008} and therefore become negligible only at distances beyond ${\sim}\SI{100}{\mega\parsec}$. Although the peculiar motions of individual galaxies always remain as a factor of uncertainty on redshift-dependent distances, it is common to model the bulk motions of galaxy groups in the nearby Universe to reduce systematic errors. To this end, different flow models with different levels of complexity have been constructed. They range from a simple correction for inflow onto the Virgo cluster \citep[e.g.,][]{Paturel1997} to a modeling of galaxy streamlines and basins of attraction out to a systemic velocity of ${\sim}\SI{30000}{\kilo\meter\per\second}$, as in the case of the CF4 flow model \citep{Courtois2023, Dupuy2023, Valade2024}.

One remaining question is how to assign realistic uncertainties to such flow distances\footnote{We use the general term "flow distance" to refer to both distances from the basic Hubble law and from flow models that account for peculiar motions.}. These uncertainties are of crucial importance because they will propagate into the uncertainties of derived quantities, for example masses, sizes, and star formation rates, as well as into the measured scatter of scaling relations like the BTFR \citep{Lelli2016b} or the FP \citep{Cappellari2011}. Better measurements of this intrinsic scatter could put tight constraints on models of dark matter and modified dynamics, but they require a very good understanding of the uncertainties involved \citep{Lelli2022}.

To date, different authors have dealt differently with the estimation of flow distance uncertainties. For example, for flow distances in the Surface Photometry and Accurate Rotation Curves (SPARC) database \citep{Lelli2016a}, a scheme is employed in which uncertainties decrease in different distance bins out to \SI{80}{\mega\parsec}, reflecting the declining influence of peculiar velocities at larger distances. Similarly, \citet{Jones2018} estimated the statistical uncertainty on flow distances as the root sum squared of the uncertainty of the Hubble constant $H_0$ and the uncertainty introduced by a constant peculiar velocity with magnitude \SI{160}{\kilo\meter\per\second}. A more data-driven approach was employed for the galaxies in the ATLAS\textsuperscript{3D} sample \citep{Cappellari2011}. These authors used Hubble distances with a correction for Virgocentric inflow for a subset of their database without other distance indicators. To estimate uncertainties, they correlated these Hubble distances with the NASA/IPAC Extragalactic Database (NED) primary distances \citep{Steer2017}. This gave them a common sample of 285 galaxies when a minimum of three independent distance measurements from the NED distance compendium was required and 692 galaxies without this requirement. The uncertainty was calculated as the root mean square (rms) error around the one-to-one relation, which resulted in ${\sim}21\%$ and ${\sim}27\%$ for the two samples, respectively. Furthermore, they found that only the inclusion of Virgocentric inflow significantly reduced the scatter around the one-to-one relation, while the inclusion of other attractors did not. However, because of their limited sample, which reaches maximum distances of only ${\sim}\SI{50}{\mega\parsec}$, these authors were unable to estimate the expected distance-dependence of the uncertainties.

In this study, we followed a similar method as \citet{Cappellari2011} to estimate uncertainties on flow distances empirically. To this aim, we compared flow distances to primary distances from the CF4 distance database. Thanks to the large sample size of this database, we were able to calculate the uncertainties in different distance bins. This allowed us to derive a functional expression for calculating the uncertainty on a flow distance measurement based on the flow distance itself.

The remainder of the paper is structured as follows. In Sect.~\ref{SecData}, we describe our sample of primary distances, introduce the four types of flow distances for which we determine uncertainties, and compare said flow distances to each other. Section~\ref{SecVZoI} describes our treatment of the Virgo cluster, which introduces large uncertainties due to peculiar motions, while Sect.~\ref{SecUnc} gives our final uncertainty scheme for all flow distances considered. We conclude in Sect.~\ref{SecSummary}.

\section{Data}  \label{SecData}

\subsection{Primary distances}  \label{SecDataPrimary}

Our starting point is the CF4 database \citep{Tully2023}, which collects distances to 55874 galaxies out to ${\sim}\SI{500}{\mega\parsec}$ and can be accessed as part of the Extragalactic Distance Database (EDD; \citeauthor{Tully2009} \citeyear{Tully2009})\footnote{http://edd.ifa.hawaii.edu}. The distances were derived with eight different methodologies (in order of increasing average uncertainties): the TRGB \citep{Lee1993}, the CPLR \citep{Leavitt1912}, surface brightness fluctuations (SBF; \citeauthor{Tonry1988} \citeyear{Tonry1988}), SNIa \citep{Phillips1993}, masers \citep{Humphreys2013, Reid2019, Pesce2020}, SNII \citep{Hamuy2002}, the TFR or BTFR \citep{Tully1977, McGaugh2000, Lelli2019}, and the FP \citep{Dressler1987, Djorgovski1987}. The distance methodologies are placed on a common distance scale by membership in $38057$ galaxy groups (of which roughly $32000$ have only one member). The zero-point calibration is provided by the geometrical maser distances, the TRGB, and the CPLR. The latter two are anchored in the geometrical maser distance to NGC~4258 \citep{Reid2019} and in stellar parallaxes, with the CPLR using an additional eclipsing binary distance to the Large Magellanic Cloud \citep{Pietrzynski2019}.

In future studies, we will use distances and their associated uncertainties to investigate, among other things, the dynamical laws of galaxies \citep{Lelli2022}. Because of this, we excluded distances based on kinematic methods, namely the TFR, BTFR, and FP. These methods are by far the largest contributors to the CF4 distances, but also the ones with the largest uncertainties ($20-25\%$; \citeauthor{Tully2023} \citeyear{Tully2023}). Consequently, we obtain a subsample of only $1956$ instead of $55874$ galaxies, which we label "Cosmicflows-4 High Quality" (\mbox{CF4-HQ}). As we show in Appendix~\ref{AppKinematic}, the main effect of keeping these kinematic distances would be a strong increase of distance uncertainties, which could therefore no longer be assumed to trace the uncertainties of flow distances.

For galaxies with distances from multiple methods, we did not use the averages provided by the CF4 team. Instead, we chose between the different methodologies according to the following priority scheme, which reflects the uncertainties of each method (in brackets; \citeauthor{Tully2023} \citeyear{Tully2023}): 1. TRGB ($5\%$), 2. CPLR ($5\%$), 3. SBF ($5\%$), 4. SNIa ($7\%$), 5. masers ($10\%$), and 6. SNII ($15\%$). TRGB and CPLR distances have similar uncertainties of the order of $5\%$. TRGB distances are available for $446$ galaxies, while CPLR distances are available for $69$ galaxies, so we gave priority to the former for the sake of consistency. Furthermore, we prioritized CPLR over SBF due to potential systematics for late-type galaxies, for which SBF are more uncertain than for early-type galaxies \citep{Tonry1988, Greco2021}.

\begin{figure}
    \centering
    \includegraphics[width = \columnwidth]{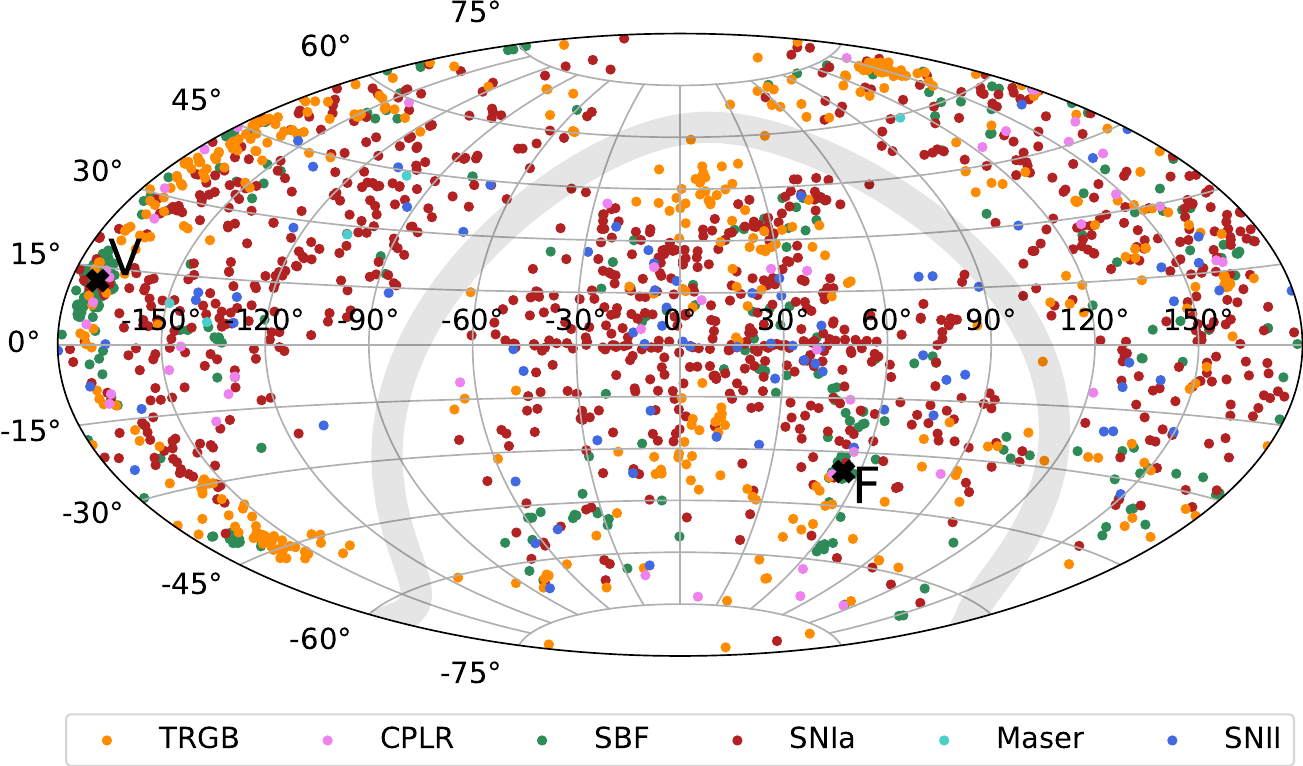}
    \caption{All-sky distribution in equatorial coordinates of the $1956$ \mbox{CF4-HQ} galaxies described in Sect.~\ref{SecDataPrimary}. The galaxies are color-coded by our preferred primary distance methodologies. The gray band indicates the disk of the Milky Way. The black crosses annotated with "V" and "F" mark the positions of the Virgo and the Fornax cluster, respectively.}
    \label{FigAllSkyDistribution}
\end{figure}

Figure~\ref{FigAllSkyDistribution} shows the all-sky distribution of the $1956$ \mbox{CF4-HQ} galaxies. The sky is well sampled with the exception of the zone of avoidance, which is expected due to strong dust obscuration in the plane of the Milky Way. The figure also highlights the positions of the Virgo and Fornax clusters, which were the targets of dedicated SBF programs with the \textit{Hubble Space Telescope} Advanced Camera for Surveys \citep{Mei2007, Blakeslee2009, Blaskeslee2010} and the Canada-France-Hawaii Telescope \citep{Cantiello2018}.

\begin{figure*}
    \centering
    \includegraphics[width = \textwidth]{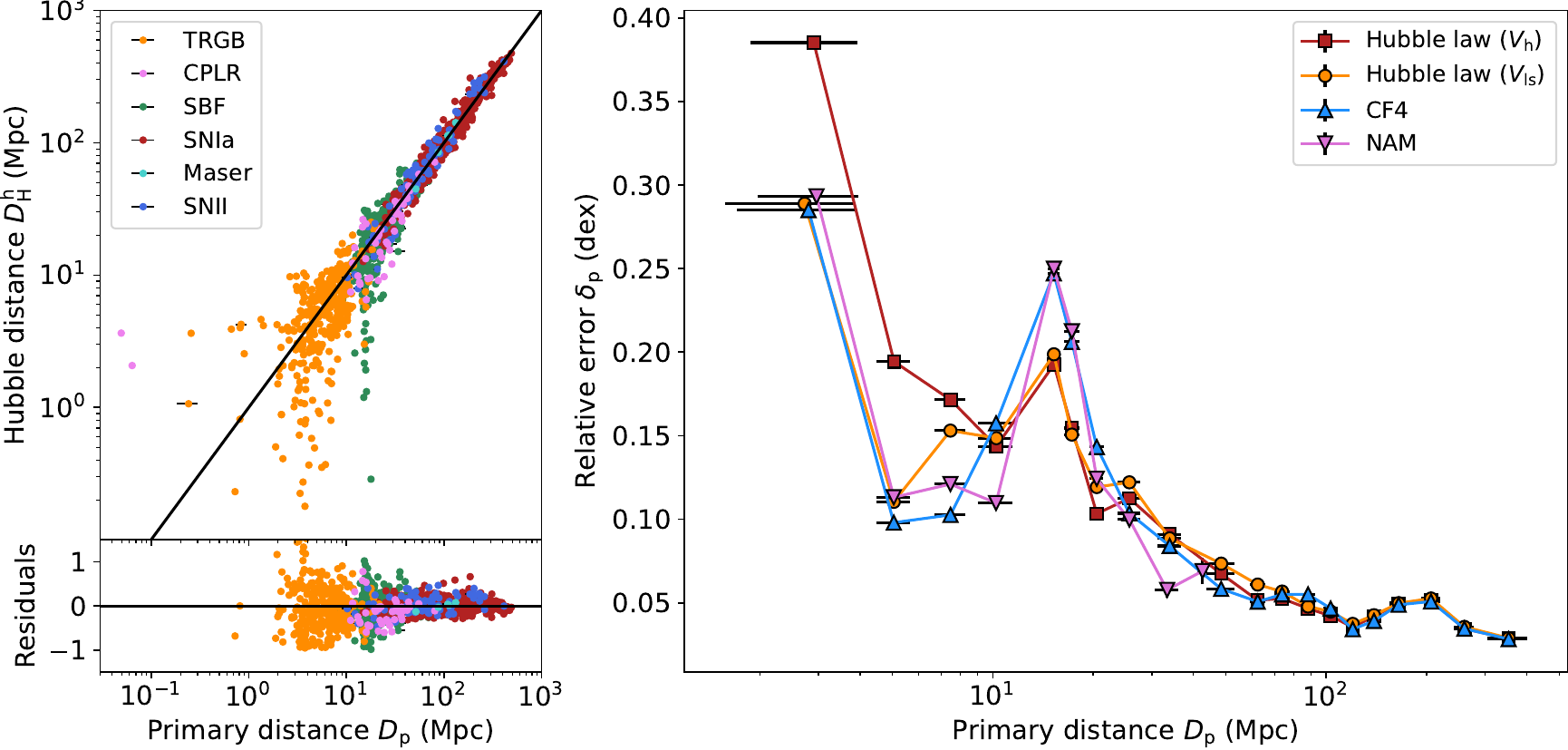}
    \caption{Heliocentric Hubble distances and relative uncertainties against the primary distances of \mbox{CF4-HQ} galaxies. \textit{Left panel:} Heliocentric Hubble distances $\mathrm{D_\mathrm{H}^\mathrm{h}}$ versus primary distances $D_\mathrm{p}$ for $1866$ galaxies from \mbox{CF4-HQ}. The galaxies are color-coded by the primary distance methodology as given in the legend. The solid black line is the one-to-one relation. The bottom panels show the residuals around the one-to-one line calculated as $(D_\mathrm{H}^\mathrm{h} - D_\mathrm{p})/D_\mathrm{p}$. We restricted the y-range to $[-1.5, +1.5]$. \textit{Right panel:} Relative error on heliocentric and local-sheet Hubble distances, CF4 distances, and NAM distances versus primary distance. See Sect.~\ref{SecPrel} for details.}
    \label{FigHubbleVsPrimary}
\end{figure*}

\subsection{Flow distances}

Our goal was to construct uncertainty schemes for distances derived with the following four flow methodologies:
\begin{enumerate}
    \item Hubble distances $D_\mathrm{H}^\mathrm{h}$ in the heliocentric reference frame are derived from the Hubble law $D_\mathrm{H}^\mathrm{h} = f(z_\mathrm{h})V_\mathrm{h}/H_0$, with $V_\mathrm{h}$ the heliocentric velocity of a galaxy, $z_\mathrm{h}$ the heliocentric redshift, $H_0$ the local Hubble parameter, and $f(z_\mathrm{h})$ a cosmological correction due to the accelerated expansion of the Universe, which is given explicitly below.
    \item Hubble distances $D_\mathrm{H}^\mathrm{ls}$ in the local-sheet reference frame are derived via $D_\mathrm{H}^\mathrm{ls} = f(z_\mathrm{ls})V_\mathrm{ls}/H_0$, with $V_\mathrm{ls}$ the local sheet velocity of a galaxy and $z_\mathrm{ls}$ the corresponding redshift. The local sheet frame was defined by \citet{Tully2008} as the velocity frame that minimizes the peculiar velocities of galaxies with primary distances between $1.1$ and \SI{7}{\mega\parsec}. Therefore, the local-sheet frame acts as a sort-of first-order flow correction.
    \item CF4 distances $D_\mathrm{CF4}$ are derived with the CF4 flow model available as part of the EDD. The CF4 model reconstructs the 3D density and velocity fields in the linear regime from the distances and peculiar velocities of galaxies in the CF4 catalog \citep{Courtois2023, Dupuy2023, Valade2024}. The velocity fields can then be used to calculate the distances of galaxies up to \SI{500}{\mega\parsec} from their systemic velocities, preferentially in the local-sheet frame.
    \item Interpolated numerical action methods (NAM) distances $D_\mathrm{NAM}$ are derived with the NAM flow model, also available as part of the EDD. The underlying NAM calculation reconstructs gravitationally induced trajectories from the distances and masses of Cosmicflows-3 galaxies \citep{Tully2016} in a nonlinear fashion \citep{Shaya2017}, while the implementation used here interpolates between grid points derived with this calculation \citep{Kourkchi2020}. Distances up to \SI{38}{\mega\parsec} can be calculated from the systemic velocity of a galaxy, preferentially using the galactic-standard-of-rest (GSR) velocity $V_\mathrm{gsr}$.
\end{enumerate}
Throughout the paper, we use $H_0 = \SI{75}{\kilo\meter\per\second\per\mega\parsec}$, from the CF4 team \citep{Tully2023}. This value is basically the same as that given by the normalization of the BTFR \citep{Schombert2020}.

The CF4 catalog provides radial velocities in the cosmic microwave background (CMB) frame $V_\mathrm{CMB}$. For the galaxies in \mbox{CF4-HQ}, we converted $V_\mathrm{CMB}$ to $V_\mathrm{h}$ using the NED Velocity Conversion Calculator\footnote{https://ned.ipac.caltech.edu/help/velc\_help.html} with CMB parameters \citep{Fixsen1996}. Based on these heliocentric velocities, we calculated the local-sheet velocities $V_\mathrm{ls}$ necessary to compute $D_\mathrm{H}^\mathrm{ls}$ and $D_\mathrm{CF4}$ according to
\begin{equation}
    V_\mathrm{ls} = V_\mathrm{h} - 26\cos{l}\cos{b} + 317\sin{l}\cos{b} - 8\sin{b}
\end{equation}
and the GSR velocities necessary for $D_\mathrm{NAM}$ according to
\begin{equation}
    V_\mathrm{gsr} = V_\mathrm{h} + 11.1\cos{l}\cos{b} + 251\sin{l}\cos{b} + 7.25\sin{b},
\end{equation}
where $l$ and $b$ are the Galactic longitude and latitude, respectively \citep{Kourkchi2020}. For the Hubble distances, we multiplied distances from the linear Hubble law with the cosmological correction
\begin{equation}    \label{EqCosmCorr}
    f(z) = 1+\frac{1}{2}(1 - q_0)z - \frac{1}{6}(2 - q_0 - 3q_0^2)z^2,
\end{equation}
where $z$ is the redshift in the appropriate reference frame and $q_0 = -0.595$ is the deceleration parameter \citep{Wright2006, Kourkchi2020}. This correction also includes the conversion of proper distances from the linear Hubble law to luminosity distances, appropriate for all distance methodologies in the \mbox{CF4-HQ} sample except for geometric maser distances. Since we have only five of these, we did not treat them differently. The cosmological correction reaches the level of a few percent at a distance of \SI{100}{\mega\parsec}. 

Finally, for each of the four flow distance methodologies, we excluded galaxies where the method failed. For the Hubble law, this is the case when $V_\mathrm{h} < 0$ or $V_\mathrm{ls} < 0 $, giving unphysical negative distances. The CF4 and NAM flow distance calculations can fail for different reasons, in particular if a galaxy has a systemic velocity that is not represented in the flow model's distance-velocity curve in that region of the sky. This can happen due to the finite resolution of the flow models, which are constructed based on the motions of galaxy groups and clusters, but do not trace the motions of individual galaxies within these structures. In addition, CF4 has a resolution limit of ${\sim}\SI{5}{\mega\parsec}$ due to the application of linear theory \citep{Courtois2023, Dupuy2023, Valade2024}. Such failures happened mostly for galaxies in the Local Group and in the nearby M81 and Centaurus~A galaxy groups \citep{Karachentsev2005}, probably because the large peculiar velocities in these groups are below the resolution limits of the models. The exclusion of galaxies where the flow methodologies fail reduces the \mbox{CF4-HQ} sample of $1956$ galaxies to $1866$ for $D_\mathrm{H}^\mathrm{h}$, $1917$ for $D_\mathrm{H}^\mathrm{ls}$, and $1881$ for $D_\mathrm{CF4}$. In the case of NAM, the flow calculator is also limited to $D_\mathrm{NAM} < \SI{38}{\mega\parsec}$, so the sample is reduced more drastically to $851$ galaxies.

\subsection{Preliminary considerations} \label{SecPrel}

In the top-left panel of Fig.~\ref{FigHubbleVsPrimary}, we show the heliocentric Hubble distances $D_\mathrm{H}^\mathrm{h}$ in our sample against the primary distances $D_\mathrm{p}$. Apart from the cosmological correction on $D_\mathrm{H}^\mathrm{h}$ from Eq.~\ref{EqCosmCorr}, this corresponds to the Hubble diagram, which plots $V_\mathrm{h}$ versus $D_\mathrm{p}$. The color-coding by distance methodology shows that the TRGB, SBF, and SNIa are the main contributors for progressively larger distances. The scatter around the one-to-one relation decreases with distance, but in a non-monotonous form. The scatter in the Hubble distances is particularly large below $D_\mathrm{p}\sim\SI{5}{\mega\parsec}$ and around the Virgo cluster at $D_\mathrm{p}\sim\SI{16.5}{\mega\parsec}$ \citep{Mei2007} and the Fornax cluster at $D_\mathrm{p}\sim\SI{20.0}{\mega\parsec}$ \citep{Blakeslee2009}, where $D_\mathrm{H}^\mathrm{h}$ can become smaller than \SI{1}{\mega\parsec}. The bottom panel in Fig.~\ref{FigHubbleVsPrimary} shows the residuals around the one-to-one relation given by $(D_\mathrm{H}^\mathrm{h} - D_\mathrm{p})/D_\mathrm{p}$. In particular, the residuals can be as large as $100\%$ below \SI{5}{\mega\parsec} and around the distance of the Virgo cluster. The residuals are even larger for galaxies with $D_\mathrm{p} < \SI{1}{\mega\parsec}$, which reside in the Local Group (not shown in the plot).

Next, we calculated the relative error on the flow distances, that is, the normalized rms error around the one-to-one relation, in ten equally spaced quantiles of primary distance, therefore capturing $0 - 10\%$, $10 - 20\%$, and so on of the sample. This was done according to
\begin{equation}    \label{EqSD}
    \delta_\mathrm{p/f}\,[\SI{}{\dex}] = \frac{1}{\bar{D}_\mathrm{p/f}\ln{10}}\sqrt{\frac{\sum_{i = 1}^N(D_\mathrm{f,i} - D_\mathrm{p,i})^2}{N}},
\end{equation}
where $N$ is the number of galaxies in quantile $i$, $\bar{D}_\mathrm{p/f}$ is either the mean primary or the mean flow distance in the quantile, and $D_\mathrm{f}\in[D_\mathrm{H}^\mathrm{h}, D_\mathrm{H}^\mathrm{ls}, D_\mathrm{CF4}, D_\mathrm{NAM}]$ is the flow model distance. The factor $1/\ln{10}$ converts the relative error to units of \SI{}{\dex}. For the CF4 and NAM flow models, multiple distance solutions may be possible for a given systemic velocity in some particular sky region, especially around massive clusters like Virgo or Fornax. Throughout the paper, we always use the flow model distance closest to the primary distance of a galaxy. In addition to the rms error in Eq.~\ref{EqSD}, we also tried using median absolute deviations $\mathrm{MAD} = \mathrm{median}(|X - \tilde{X}|)$, where $\tilde{X}$ is the median of the data $X$. The MAD relative errors are always smaller than the rms relative errors, even after multiplying the MAD by $\sqrt{\pi/2}$ to obtain Gaussian-like standard deviations. This indicates that the error structure of the data is not Gaussian and contains a significant number of outliers. To be conservative, we use the rms errors from Eq.~\ref{EqSD} in the rest of the paper.

\begin{figure*}[h!]
    \centering
    \includegraphics[width = \textwidth]{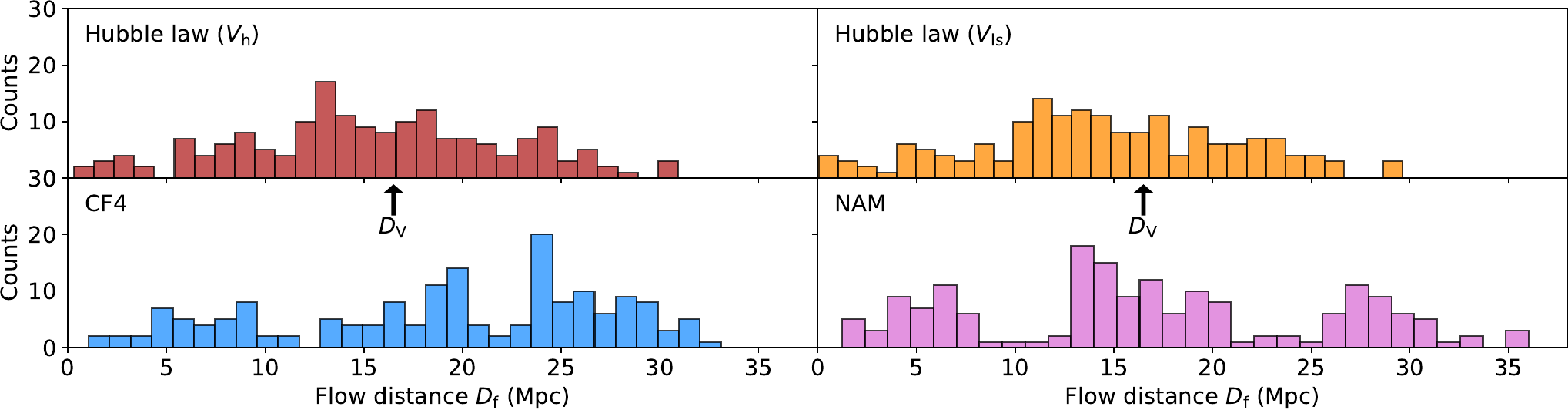}
    \caption{Flow distance histogram of galaxies within a primary distance range from $14$ to \SI{19}{\mega\parsec}, that is, around the distance of the Virgo cluster. The different panels correspond to the four types of flow distances as indicated. The black arrows annotated with $D_\mathrm{V}$ indicate the distance of the Virgo cluster at \SI{16.5}{\mega\parsec}.}
    \label{FigVirgoHistograms}
\end{figure*}

The right-hand panel of Fig.~\ref{FigHubbleVsPrimary} shows the mean relative error $\delta_\mathrm{p}$ against the primary distance $D_\mathrm{p}$ for the four types of flow distances. The data is binned into ten equally spaced quantiles in $D_\mathrm{p}$. We used the quantile limits determined from the subsample of the heliocentric Hubble law for all four flow methodologies to ease the comparison. This results in a small bin of only six galaxies for the last data point of the NAM distances, which should therefore be interpreted with caution. The bins are centered on the mean distance within each quantile and the horizontal error bars represent the corresponding standard deviation. Here and in the rest of the paper, the uncertainties on the derived relative error $\delta_\mathrm{p/f}$ are calculated via Gaussian propagation of the error on $D_\mathrm{p}$. They are typically smaller than the symbols.

The overall trend for all four flow methodologies is a decline in relative error from ${\sim}0.3-\SI{0.4}{\dex}$ at \SI{3}{\mega\parsec} to ${\sim}\SI{0.05}{\dex}$ for distances larger than \SI{100}{\mega\parsec}. This is due to the declining relative contribution of peculiar velocities to redshifts as distances increase. In all four cases, the general trend is broken by an increase in relative error to ${\sim}0.20-\SI{0.25}{\dex}$ between $10$ and \SI{20}{\mega\parsec}. This peak in relative error can be removed by excluding galaxies around the primary distances and angular positions of the Virgo cluster and the Fornax cluster. Therefore, we attribute it to large peculiar motions inside and in the vicinities of these two cluster.

Comparing the different flow distances, the Hubble law with local-sheet velocities is almost identical to that with heliocentric velocities, except for primary distances lower than \SI{10}{\mega\parsec}, where it offers a significant improvement of up to \SI{0.1}{\dex}. This confirms that the local-sheet frame is indeed less biased by the motion of the Sun and the Milky Way with respect to the Hubble flow. The CF4 and NAM models improve over the heliocentric Hubble distances below \SI{5}{\mega\parsec}, but are comparable to the local-sheet Hubble distances. This is likely because they have effective resolutions of a few megaparsec, so they cannot model the motions of individual galaxies inside nearby groups. However, these two flow models offer an additional improvement up to distances of roughly \SI{10}{\mega\parsec}, where the effect of the Virgo and Fornax clusters becomes important. Interestingly, the CF4 and NAM flow models have relative errors larger by about \SI{0.05}{\dex} around these clusters compared to the Hubble distances. For distances larger than ${\sim}\SI{20}{\mega\parsec}$, the two flow models have smaller relative errors than the Hubble distances. Beyond ${\sim}\SI{38}{\mega\parsec}$, the NAM model stops being applicable, while the CF4 model becomes indistinguishable from the two implementations of the Hubble law beyond \SI{100}{\mega\parsec}. At these distances, the relative errors are dominated by the uncertainty in the value of $H_0$ rather than by the use of a specific flow model.

We conclude that the general distance-dependent trend in the relative errors is only interrupted by peculiar motions inside and around the Virgo cluster and to a lesser extent the Fornax cluster. This affects all four methodologies. Consequently, we treat the 3D volume in distance and separation around Virgo differently from the rest of the sky, as we describe in the next section. In Appendix~\ref{AppFornax}, we describe how a similar treatment of the Fornax cluster has a negligible effect on our results and is therefore not a part of our uncertainty scheme.

\section{The Virgo Zone of Influence}   \label{SecVZoI}

With the term "Virgo Zone of Influence" (Virgo ZoI), we refer to the three-dimensional region around the Virgo cluster where we assign uncertainties with a different scheme compared to the rest of the sky.
For defining the Virgo ZoI, we first investigated the range of flow distances spanned by Virgo, and then the range in angular separation from Virgo where the scatter is increased with respect to the overall trend at the same flow distances. This requires knowledge of the actual extent of Virgo along the line of sight. The distance to Virgo is $16.5\pm0.1\pm1.1\,$\SI{}{\mega\parsec} (statistical and systematic uncertainties) and its $2\sigma$ back-to-front line-of-sight depth is \SI{2.4\pm0.4}{\mega\parsec} \citep{Mei2007}. To be conservative, we used twice this range for the primary distance extent of the Virgo ZoI, that is, $14$ to \SI{19}{\mega\parsec}. While the Virgo cluster has additional structure on its far side, in particular the W~cloud at two times the distance of the cluster itself \citep{deVaucouleur1961, Binggeli1993}, the inclusion of this structure in the primary distance range has no significant effect on our results. In addition, our adopted range has the advantage of not including the Fornax cluster around \SI{20}{\mega\parsec}. 

In Fig.~\ref{FigVirgoHistograms}, we show the histograms of flow distances $D_\mathrm{f}$ for galaxies with a primary distance in the range from $14$ to \SI{19}{\mega\parsec}. Galaxies with primary distances around Virgo can have flow distances that are as low as \SI{1}{\mega\parsec} and as large as \SI{30}{\mega\parsec}. The number of galaxies in the four histograms is roughly the same, with $176$, $175$, $169$, and $173$ for the heliocentric Hubble law, the local-sheet Hubble law, CF4, and NAM, respectively. Regarding our definition of the Virgo ZoI, since galaxies at the edges of these histograms contribute most strongly to the errors of flow distances, we use the full range of the histograms. For simplicity, we assume the same range for all four types of flow distances for the Virgo ZoI, namely $D_\mathrm{f}\in(\SI{1}{\mega\parsec}, \SI{33}{\mega\parsec})$.

\begin{figure}
    \centering
    \includegraphics[width = \columnwidth]{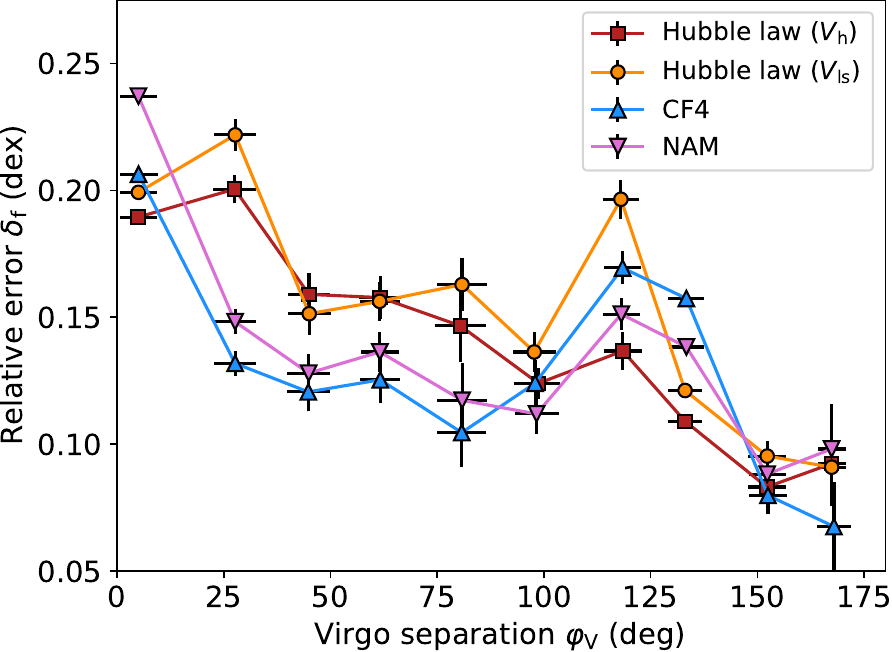}
    \caption{Relative error for galaxies with flow distances between $1$ and \SI{33}{\mega\parsec} against the angular separation from the Virgo cluster. The four curves correspond to the four flow methodologies as given in the legend.}
    \label{FigSeparation}
\end{figure}

Next, we determined the range of the Virgo ZoI in right ascension and declination. Figure~\ref{FigSeparation} shows the relative error on the four different flow distances versus the angular separation from the Virgo cluster $\varphi_\mathrm{V}$ for galaxies in the flow distance range $(\SI{1}{\mega\parsec}, \SI{33}{\mega\parsec})$. The data is grouped into ten equally spaced bins from $0$ to \SI{180}{\deg}. For the center of Virgo, we assume $(\alpha, \delta) = (\SI{187.71}{\deg}, \SI{12.39}{\deg})$, which is the position of M87. The bin centers are given by the mean separations of the galaxies in the respective bins and the horizontal error bars by the corresponding standard deviations.

All four flow models show a similar general trend. The relative error is at maximum near the Virgo cluster, with values between ${\sim}0.2$ and \SI{0.25}{\dex}. For larger separations from Virgo, the relative error decreases and reaches a roughly constant level between $0.1$ and \SI{0.17}{\dex} for Virgo separations up to \SI{100}{\deg}. Between $100$ and \SI{150}{\deg}, an additional peak is present around the separation of the Fornax cluster from Virgo. This is likely due to peculiar motions around Fornax. Around the Virgo peak, the drop-off is steeper for CF4 and NAM, which reach the base level already at $\varphi_\mathrm{V}\sim\SI{30}{\deg}$ compared to $\varphi_\mathrm{V}\sim\SI{45}{\deg}$ the two implementations of the Hubble law. This indicates that the CF4 and NAM models might successfully model peculiar motions in the vicinity of the Virgo cluster, even though they cannot model them inside the cluster. Following this, we adopt a different angular extent $\varphi_\mathrm{VZoI}$ for the Virgo ZoI depending on the type of flow distance, namely
\begin{equation}
    D_\mathrm{f}\in(\SI{1}{\mega\parsec}, \SI{33}{\mega\parsec}) \hspace{.5cm} \mathrm{and}\hspace{.5cm} \varphi_\mathrm{Virgo} < \SI{45}{\deg}
\end{equation}
for the two implementations of the Hubble law and
\begin{equation}
    D_\mathrm{f}\in(\SI{1}{\mega\parsec}, \SI{33}{\mega\parsec}) \hspace{.5cm} \mathrm{and}\hspace{.5cm} \varphi_\mathrm{Virgo} < \SI{30}{\deg}
\end{equation}
for the CF4 and NAM flow models.

We did not try to model the trends of the relative error with $\varphi_\mathrm{V}$ inside the Virgo ZoI. Instead, we calculated the mean relative error $\delta_\mathrm{VZoI}$ in the first two $\varphi_\mathrm{V}$-bins for each flow distance method, which we conservatively assign to each galaxy located inside the Virgo ZoI. The resulting relative errors $\delta_\mathrm{VZoI}$ are \SI{0.19}{\dex}, \SI{0.21}{\dex}, \SI{0.17}{\dex}, and \SI{0.19}{\dex} for $D_\mathrm{H}^\mathrm{h}$, $D_\mathrm{H}^\mathrm{ls}$, $D_\mathrm{CF4}$, and $D_\mathrm{NAM}$, respectively.

\begin{figure*}
    \centering
    \includegraphics[width = \textwidth]{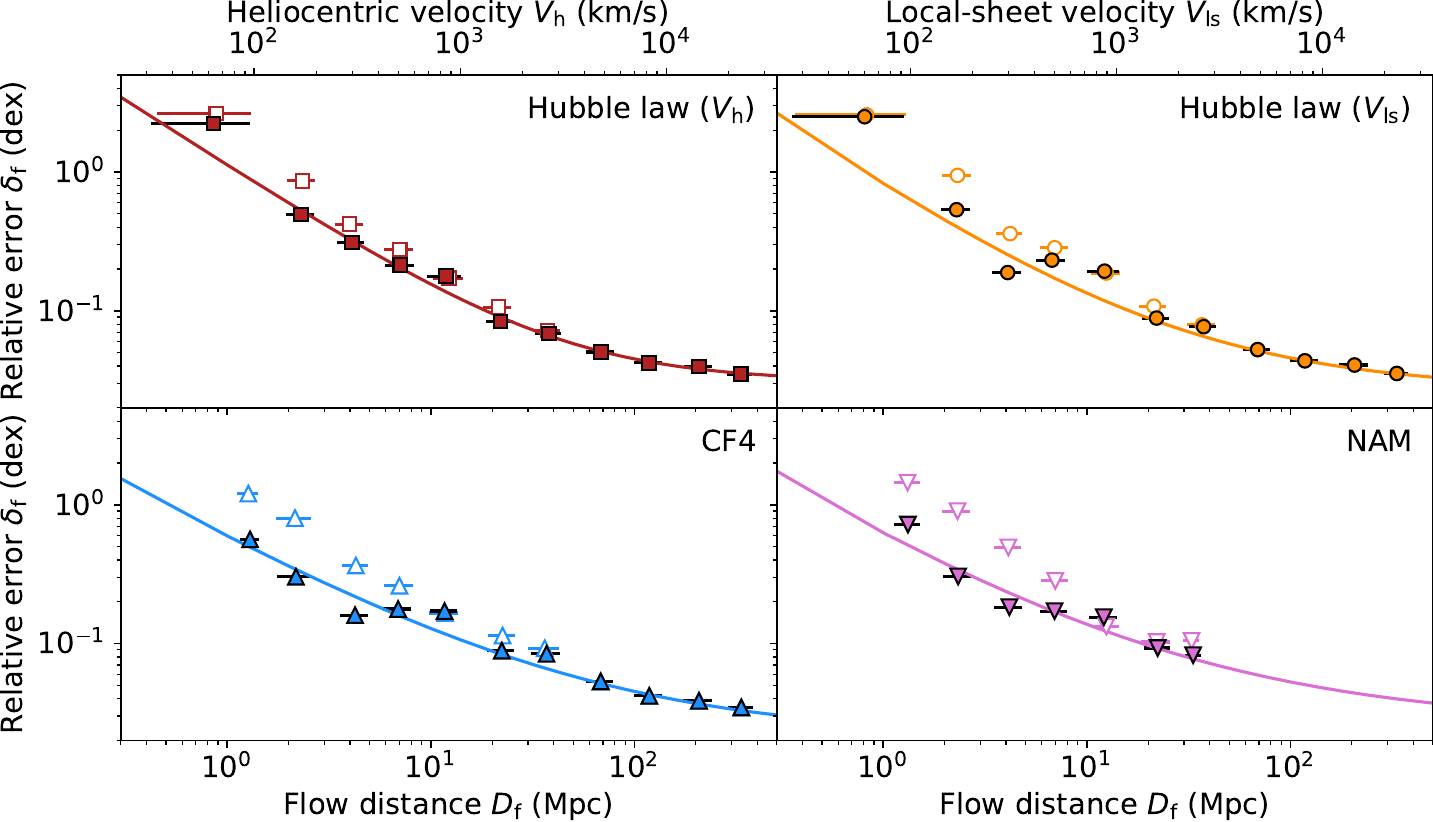}
    \caption{Relative error versus flow distance excluding (filled symbols) and including (open symbols) galaxies in the Virgo ZoI for the four flow methodologies. For the two implementations of the Hubble law, the upper axes show the corresponding systemic velocities. The solid curves are the best-fit power laws from Eq.~\ref{EqDBPL} to the data excluding the Virgo ZoI.}
    \label{FigSDExcVirgo}
\end{figure*}

\section{Uncertainties on flow distances} \label{SecUnc}

Outside of the Virgo ZoI, our goal was to track the dependence of the relative errors on the flow distance of a galaxy. In other words, we wanted to define a function $\delta_\mathrm{f}(D_\mathrm{f})$ that allows for the quick calculation of the uncertainty of a galaxy distance determined with one of the four flow methodologies. In Fig.~\ref{FigSDExcVirgo}, we show $\delta_\mathrm{f}$ against the flow distance $D_\mathrm{f}$ for all four methodologies, both excluding and including the Virgo ZoI. The exclusion reduces the \mbox{CF4-HQ} sample sizes from $1866$ to $1489$ for $D_\mathrm{H}^\mathrm{h}$, from $1917$ to $1543$ for $D_\mathrm{H}^\mathrm{ls}$, from $1881$ to $1634$ for $D_\mathrm{CF4}$, and from $851$ to $602$ for $D_\mathrm{NAM}$. All four flow distances use the same bins, namely $20$ logarithmic bins between $0.01$ and \SI{480}{\mega\parsec} in $D_\mathrm{f}$. This range is large enough to not exclude any galaxies. Of these bins, the first ten are merged to ensure a minimum number of $20$ galaxies per bin. Since the NAM model does not extend beyond a flow distance of \SI{38}{\mega\parsec}, it does not have data in the last four flow distance bins. The number of galaxies per bin ranges from $21$ to $361$, with larger bins for larger distances and most bins having more than $100$ galaxies.

In all four cases, there is a clear trend of decreasing relative error with increasing flow distance. Furthermore, the Virgo and Fornax peak, clearly visible in Fig.~\ref{FigHubbleVsPrimary}, cannot be seen in this figure. However, this is not an effect of excluding the Virgo ZoI, but of plotting versus $D_\mathrm{f}$ rather than versus $D_\mathrm{p}$. As shown in Sect.~\ref{SecVZoI}, the use of flow distances spreads the Virgo cluster out between $1$ and \SI{33}{\mega\parsec} and similarly the Fornax cluster, as shown in Appendix~\ref{AppFornax}. Instead, excluding galaxies in the Virgo ZoI serves to lower the overall relative errors below flow distances of \SI{30}{\mega\parsec}. The data for the sample excluding the Virgo ZoI is given in Table~\ref{TabSDExcVirgo}.

\begin{table}
    \caption{Relative error $\delta_\mathrm{f}$ in different flow distance bins for the four flow methodologies after removing the Virgo ZoI.}
    \label{TabSDExcVirgo}
    \centering
    \begin{tabular}{lcccc}
        \hline
        \hline
        $D_\mathrm{f}$ (\SI{}{\mega\parsec}) & Hubble ($V_\mathrm{h}$) & Hubble ($V_\mathrm{ls}$) & CF4 & NAM\\ \hline
        $0.01 - 1.65$ & $2.24$ & $2.50$ & $0.56$ & $0.72$\\
        $< 2.91$ & $0.49$ & $0.54$ & $0.30$ & $0.30$\\
        $< 5.13$ & $0.31$ & $0.19$ & $0.16$ & $0.18$\\
        $< 9.05$ & $0.21$ & $0.23$ & $0.18$ & $0.17$\\
        $< 16.0$ & $0.18$ & $0.19$ & $0.17$ & $0.16$\\
        $< 28.1$ & $0.08$ & $0.09$ & $0.09$ & $0.09$\\
        $< 49.6$ & $0.07$ & $0.08$ & $0.08$ & $0.08$\\
        $< 87.5$ & $0.05$ & $0.05$ & $0.05$ & \\
        $< 154$ & $0.04$ & $0.04$ & $0.04$ & \\
        $< 272$ & $0.04$ & $0.04$ & $0.04$ & \\
        $< 480$ & $0.04$ & $0.04$ & $0.04$ & \\
        \hline
    \end{tabular}
    \tablefoot{Data as described in the main text. Relative errors are given in units of \SI{}{\dex}. The NAM model does not extend beyond \SI{38}{\mega\parsec}.}
\end{table}

\begin{table*}[t]
    \caption{Best-fit parameters of the broken power law fits according to Eq.~\ref{EqDBPL} to the data in Fig.~\ref{FigSDExcVirgo} (excluding the Virgo ZoI; \textit{left-hand side}) and parameters defining the range of application of the Virgo ZoI and the relative uncertainty therein for each flow methodology (\textit{right-hand side}).}
    \label{TabParameters}
    \centering
    \begin{tabular}{lcccc|ccc}
        \hline
        \hline
        Flow distance method & $\delta_\infty$ (\SI{}{\dex}) & $\alpha$ & $D_\mathrm{tr}$ (\SI{}{\mega\parsec}) & $\kappa$ & $D_\mathrm{VZoI}$ (\SI{}{\mega\parsec})& $\varphi_\mathrm{VZoI}$ (\SI{}{\deg}) & $\delta_\mathrm{VZoI}$ (\SI{}{\dex}) \\ \hline
        Hubble ($V_\mathrm{h}$) & $0.031\pm0.005$ & $-0.9\pm0.3$ & $44\pm31$ & $1.0\pm0.5$ & $7.7-33$ & $45$ & $0.19$\\
        Hubble ($V_\mathrm{ls}$) & $0.027\pm0.039$ & $-1.0\pm3.0$ & $25\pm255$ & $1.5\pm4.7$ & $5.2-33$ & $45$ & $0.21$\\
        CF4 & $0.022\pm0.027$ & $-0.8\pm0.9$ & $46\pm190$ & $1.8\pm3.3$ & $6.2-33$ & $30$ & $0.17$\\
        NAM & $(0.026)$ & $-0.9\pm1.2$ & $22\pm131$ & $2.0\pm1.9$ & $5.7-33$ & $30$ & $0.19$\\
        \hline
    \end{tabular}
    \tablefoot{The ranges are $1\sigma$ confidence intervals. The $\delta_\infty$ parameter was fixed for the NAM model as described in Sect.~\ref{SecUnc}. The lower distance limit of the Virgo ZoI, $D_\mathrm{VZoI}$, is given together with the upper limit, \SI{33}{\mega\parsec}, for clarity.}
\end{table*}

We fitted the distance-dependent standard deviations excluding the Virgo ZoI with a broken power law with a softening parameter, which is shown with a solid line in Fig.~\ref{FigSDExcVirgo}. At very large distances, uncertainties due to peculiar motions should become negligible and the standard deviations should approach a constant value, reflecting the uncertainty of $H_0$. To capture this, we fixed the second slope of the broken power law to $0$. Therefore, the fitting function is
\begin{equation}    \label{EqDBPL}
    \delta_\mathrm{f} = \delta_\infty {D_\mathrm{f}}^\alpha (D_\mathrm{f}^{1/\kappa} + D_\mathrm{tr}^{1/\kappa})^{-\alpha\kappa},
\end{equation}
where $\delta_\infty$ is the asymptotic relative error at large distances, $\alpha$ is the slope at small distances, $D_\mathrm{tr}$ is the transition distance, and $\kappa$ is the softening parameter. We used the orthogonal distance regression methodology for the fit \citep{Boggs1990}. In the case of the NAM model, the model does not reach large enough distances to trace the asymptotic flattening. Therefore, we fixed the $\delta_\infty$ of NAM to \SI{0.026}{\dex}, which is the arithmetic mean of the best-fit values of this parameter for the other three models. All best-fit parameters are given in Table~\ref{TabParameters}. As expected, the fitted functions approach each other for larger distances for all four flow models, while the CF4 and the NAM model have smaller uncertainties at smaller distances.

Importantly, we defined the Virgo ZoI to reflect the increased relative uncertainties around to the Virgo cluster. However, a naive application of the Virgo ZoI will give smaller uncertainties than Eq.~\ref{EqDBPL} for galaxies with flow distances smaller than about $5$ to \SI{10}{\mega\parsec}, within which the relative errors steeply decrease. Therefore, for the actual application of the Virgo ZoI, we use a different lower flow distance limit, $D_\mathrm{ZoI}$, which for each flow methodology is given by the distance where the power law reaches the relative error $\delta_\mathrm{ZoI}$ assigned within the Virgo ZoI. The values of $D_\mathrm{ZoI}$ are given in Table~\ref{TabParameters}.

In summary, our final uncertainty scheme for flow distances $D_\mathrm{f}\in[D_\mathrm{H}^\mathrm{h}, D_\mathrm{H}^\mathrm{ls}, D_\mathrm{CF4}, D_\mathrm{NAM}]$ takes the following form. The relative error $\delta_\mathrm{f}$ of the flow distance $D_\mathrm{f}$ of a galaxy at a separation $\varphi_\mathrm{V}$ from the Virgo cluster ($\alpha = \SI{187.71}{\deg}$, $\delta = \SI{12.39}{\deg}$) is
\begin{equation}    \label{EqFinal}
    \delta_\mathrm{f}\,[\SI{}{\dex}] = \begin{cases} 
        \delta_\mathrm{VZoI}\mathrm{,\hspace{0.1cm}if}\hspace{0.1cm} \text{$D_\mathrm{f}\in(D_\mathrm{VZoI}, \SI{33}{\mega\parsec}) \hspace{0.1cm}\mathrm{and}\hspace{0.1cm} \varphi_\mathrm{V} < \varphi_\mathrm{VZoI}$}\\
    \\
        \delta_\infty D_\mathrm{f}^\alpha (D_\mathrm{f}^{1/\kappa} + D_\mathrm{tr}^{1/\kappa})^{-\alpha\kappa},\hspace{0.1cm}\text{else},
        
    \end{cases}
\end{equation}
with $\delta_\mathrm{VZoI}$ the relative error assigned inside the Virgo ZoI, $D_\mathrm{VZoI}$ the closer distance limit of the Virgo ZoI, and $\varphi_\mathrm{VZoI}$ the angular extent of the Virgo ZoI. All these parameters depend on the flow model chosen and are given explicitly in Table~\ref{TabParameters}. The same is the case for the parameters $\delta_\infty$, $\alpha$, $\kappa$, and $D_\mathrm{tr}$ of the function in the second line.

\section{Conclusion}   \label{SecSummary}

We constructed a scheme for assigning relative errors $\delta_\mathrm{f}$ to galaxy flow distances $D_\mathrm{f}$ based only on said flow distance and a galaxy's proximity to the Virgo cluster. Our uncertainty scheme is based on the comparison of galaxy flow distances to primary distances from a subsample of the CF4 database \citep{Tully2023} with TRGB, CPLR, SBF, SNIa, maser, and SNII distances, that is, excluding TFR, BTFR, and FP distances. We derived this scheme for four different flow models: 1. the Hubble law with heliocentric velocities, 2. the Hubble law with velocities in the local-sheet frame \citep{Tully2008}, 3. the CF4 flow model \citep{Courtois2023, Dupuy2023, Valade2024}, and 4. an interpolation of grid points from the NAM flow model \citep{Shaya2017, Kourkchi2020}.

The main output of this work is a simple formula, Eq.~\ref{EqFinal}, that provides the uncertainty on $D_\mathrm{f}$ as a function of $D_\mathrm{f}$. The parameters of Eq.~\ref{EqFinal} depend on the flow model chosen and are given in Table~\ref{TabParameters}. This error scheme, however, should not be used for galaxies in the surroundings of the Virgo cluster where distance uncertainties are significantly larger. Therefore, we defined a "zone of influence" of Virgo (Virgo ZoI) for each flow model. Galaxies that lie within this region should be assigned a boosted uncertainty specific to the Virgo ZoI. These boosted uncertainties, as well as the boundaries of the Virgo ZoI, are also given in Table~\ref{TabParameters}.

In the near future, we will apply our new error scheme to galaxy distances in the upcoming Broad Inventory of Galaxies with Surface Photometry and Accurate Rotation Curves (BIG-SPARC; \citeauthor{Haubner2024} \citeyear{Haubner2024}). Crucially, well-defined uncertainties on galaxy distances are important for the study of the intrinsic scatter around dynamical scaling laws \citep[e.g.,][]{Lelli2016b, Lelli2019} and to discriminate between different dark matter and modified-gravity theories \citep[e.g.,][]{Lelli2022}. For example, the controversial results about the mass distribution of ultra-diffuse galaxies were in part driven by underestimated uncertainties on flow distances \citep{Lelli2024}. More generally, having well-defined uncertainties on flow distances will be useful for a large variety of astrophysical applications, such as determining the fundamental parameters of galaxies (e.g., luminosities, masses, sizes, and SFRs) and their mutual scaling relations.

\begin{acknowledgements}
We thank the Cosmicflows team for making their calculators and datasets publicly available and for providing useful comments that benefited this paper. EDT was supported by the European Research Council (ERC) under grant agreement no.\ 101040751. This research has made use of the NASA/IPAC Extragalactic Database, which is funded by the National Aeronautics and Space Administration and operated by the California Institute of Technology.
\end{acknowledgements}

\bibliographystyle{aa}
\bibliography{references}

\begin{appendix}

\section{Kinematic primary distances}    \label{AppKinematic}

\begin{figure}[h]
    \centering
    \includegraphics[width = \columnwidth]{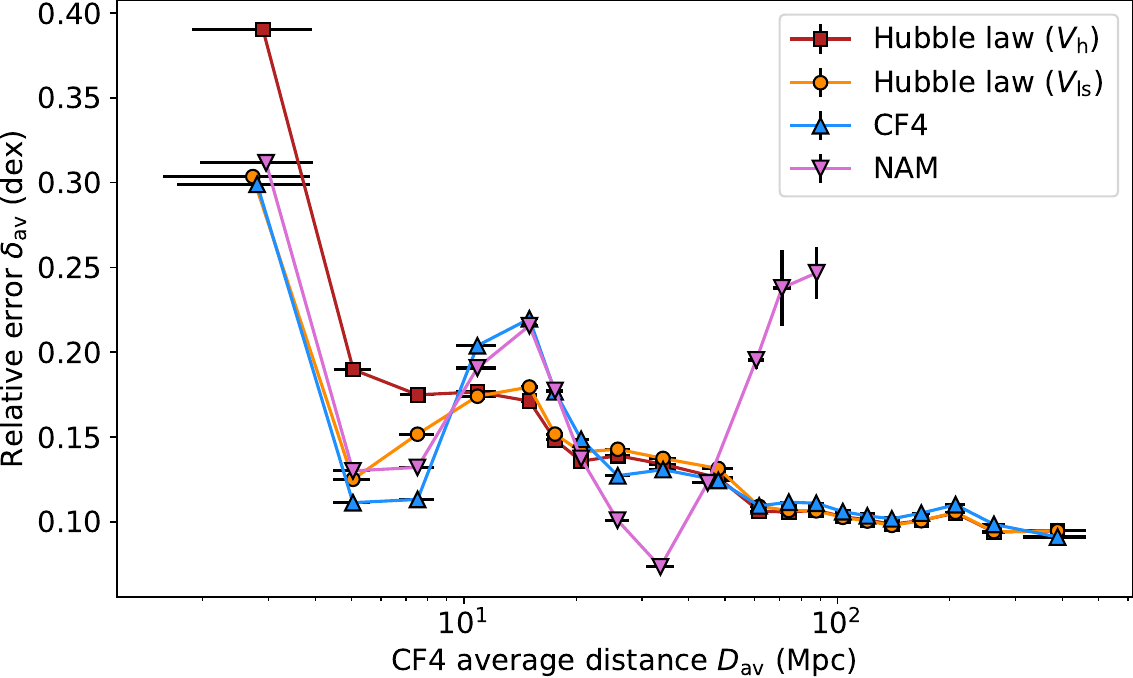}
    \caption{Relative error on heliocentric and local-sheet Hubble distances, CF4 distances, and NAM distances versus CF4 averaged distances, including TFR, BTFR, and FP. See Appendix~\ref{AppKinematic} for details.}
    \label{FigWithKinematic}
\end{figure}

Of the roughly $56000$ galaxies in the CF4 database, $54000$ have distances only from the TFR, BTFR, or the FP. We did not consider these distances in our analysis because they have significantly larger errors ($20-25\%$) than other primary distances ($5-15\%$). Still, it is worth to check their effect on our results. Figure~\ref{FigWithKinematic} is the same as the right-hand panel of Fig.~\ref{FigHubbleVsPrimary}, but using the weighted averages $D_\mathrm{av}$ of the different distance methodologies provided by the CF4 team. Importantly, the main contributions to these are BTFR and FP distances. This plot uses $55784$, $55835$, $55797$, and $2827$ galaxies for $D_\mathrm{H}^\mathrm{h}$, $D_\mathrm{H}^\mathrm{ls}$, $D_\mathrm{CF4}$, and $D_\mathrm{NAM}$, respectively. To facilitate the comparison, we fixed the bins to the same primary distance ranges as in Fig.~\ref{FigHubbleVsPrimary}.

The general trends are the same as in Fig.~\ref{FigHubbleVsPrimary}, but there are two important differences. First, the height of the Virgo peak is reduced from ${\sim}0.2$ to ${\sim}\SI{0.18}{\dex}$ for both implementations of the Hubble law and from ${\sim}0.25$ to ${\sim}\SI{0.22}{\dex}$ for CF4 and NAM. However, the peak itself is broadened. In fact, for the heliocentric Hubble law, it is not discernible as a peak at all. Second, the relative error at large distances is increased from ${\sim}0.05$ to ${\sim}\SI{0.1}{\dex}$. This means that the uncertainty of the kinematic primary distances contributes as strongly to the relative error at large distances as the uncertainties of the flow distances themselves. This violates the assumption that the relative errors calculated via Eq.~\ref{EqSD} are a good indicator of the uncertainties on flow distances, therefore giving an additional justification for the exclusion of kinematic distances from our analysis.

Furthermore, we note that the NAM model contains three additional bins at large distances, which display an unusual rising behavior. These bins contain only $21$, three, and one galaxy from lower to higher distances, compared to at least $97$ galaxies in every other bin. The relative error in these bins is very large and probably driven by outliers in the TFR, the BTFR, and the FP. In particular, since the NAM model can output a maximum flow distance of \SI{38}{\mega\parsec}, there is a selection effect to large errors for primary distances farther away than this limit. This effect is probably only noticeable for kinematic distances due to their larger uncertainty.

\newpage

\section{The Fornax cluster}    \label{AppFornax}

\begin{figure*}
    \centering
    \includegraphics[width = \textwidth]{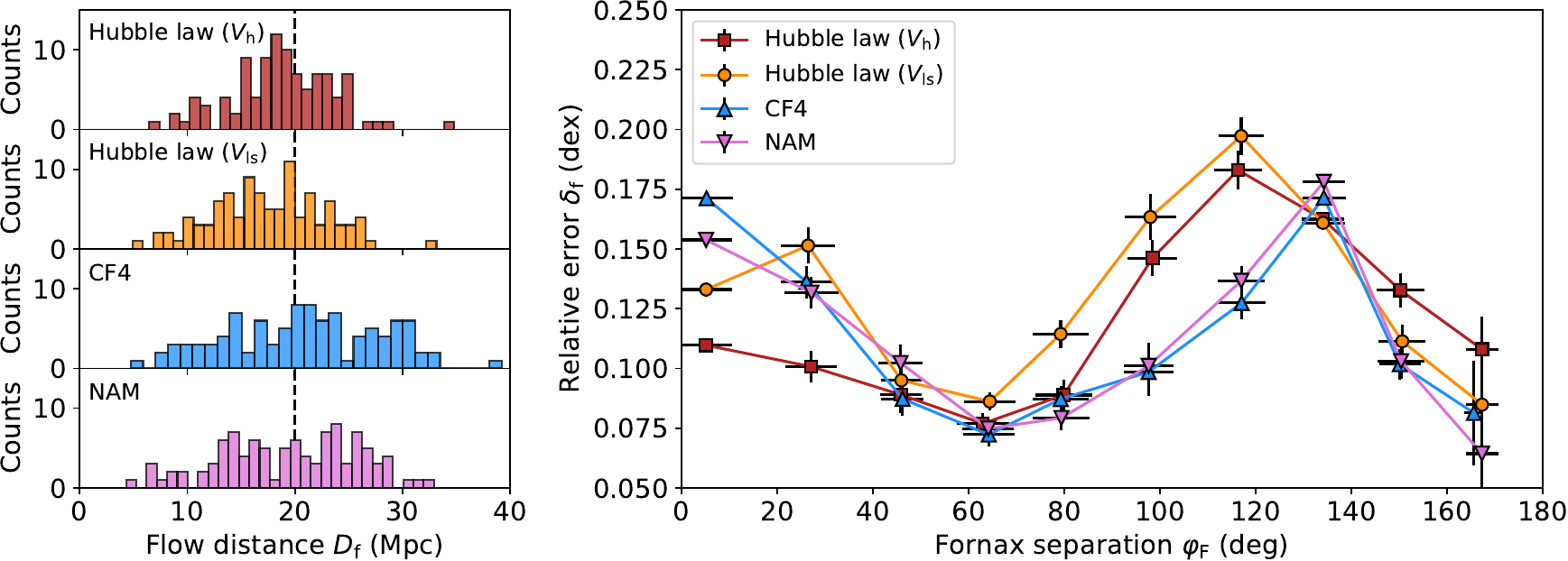}
    \caption{Flow-distance ranges and separation-dependent relative errors for the Fornax ZoI. \textit{Left panel:} Flow distance histogram of galaxies within a primary distance range from $18$ to \SI{22}{\mega\parsec}, that is, around the distance of the Fornax cluster. The different panels correspond to the four types of flow distances as indicated. The vertical dashed line marks the distance to the Fornax cluster of \SI{20}{\mega\parsec}. \textit{Right panel:} Relative error for galaxies in the distance range $(\SI{5}{\mega\parsec}, \SI{33}{\mega\parsec})$ against the angular separation from the Fornax cluster. The four curves correspond to the four flow methodologies as given in the legend. The additional peak at a separation of ${\sim}\SI{130}{\deg}$ is due to the Virgo cluster.}
    \label{FigFornaxHistograms}
\end{figure*}

The second-closest massive galaxy cluster to the Milky Way is the Fornax cluster at a distance of \SI{20}{\mega\parsec} \citep{Blakeslee2009}. Here, we checked whether peculiar motions around the Fornax cluster justify a separate treatment of uncertainties similar to the Virgo ZoI. To this end, we defined a Fornax ZoI and investigated its effect on the relative errors, following the same steps as for the Virgo ZoI. The Fornax cluster has an $2\sigma$ back-to-front line-of-sight depth of ${\sim}\SI{2}{\mega\parsec}$ \citep{Blakeslee2009}, so in analogy to the $4\sigma$ range that we used for Virgo, we employed a primary range of $(\SI{18}{\mega\parsec}, \SI{22}{\mega\parsec})$ for Fornax.

The left panel of Fig.~\ref{FigFornaxHistograms} shows the histograms of flow distances in the primary range $(\SI{18}{\mega\parsec}, \SI{22}{\mega\parsec})$ for the four respective subsamples of \mbox{CF4-HQ}. Compared to the Virgo cluster, the galaxies at the distance of the Fornax cluster span a smaller range in flow distances. For simplicity, we use the same flow distance range of $(\SI{5}{\mega\parsec}, \SI{33}{\mega\parsec})$ for the Fornax ZoI for all four flow methodologies. However, the precise choice of this range has only a negligible effect on our further analysis.

The right panel of Fig.~\ref{FigFornaxHistograms} gives the relative error of galaxies in the $(\SI{5}{\mega\parsec}, \SI{33}{\mega\parsec})$ flow distance range versus the angular separation from the Fornax cluster $\varphi_\mathrm{F}$. For the center of Fornax, we assume $(\alpha, \delta) = (\SI{54.62}{\deg}, \SI{-35.45}{\deg})$, which is the position of NGC~1399. The right panel of Fig.~\ref{FigFornaxHistograms} is very similar to Fig.~\ref{FigSeparation}. For all four types of flow distances, the relative error is larger close to Fornax and decreases for increasing separations, up to a separation of ${\sim}\SI{60}{\deg}$. Furthermore, a second, even higher and broader peak is located around a separation of \SI{130}{\deg}. This is probably due to peculiar motions around the Virgo cluster. We adopt a threshold of $\varphi < \SI{50}{\deg}$ for the Fornax ZoI, broader than that of the Virgo ZoI.

\begin{figure*}
    \centering
    \includegraphics[width = \textwidth]{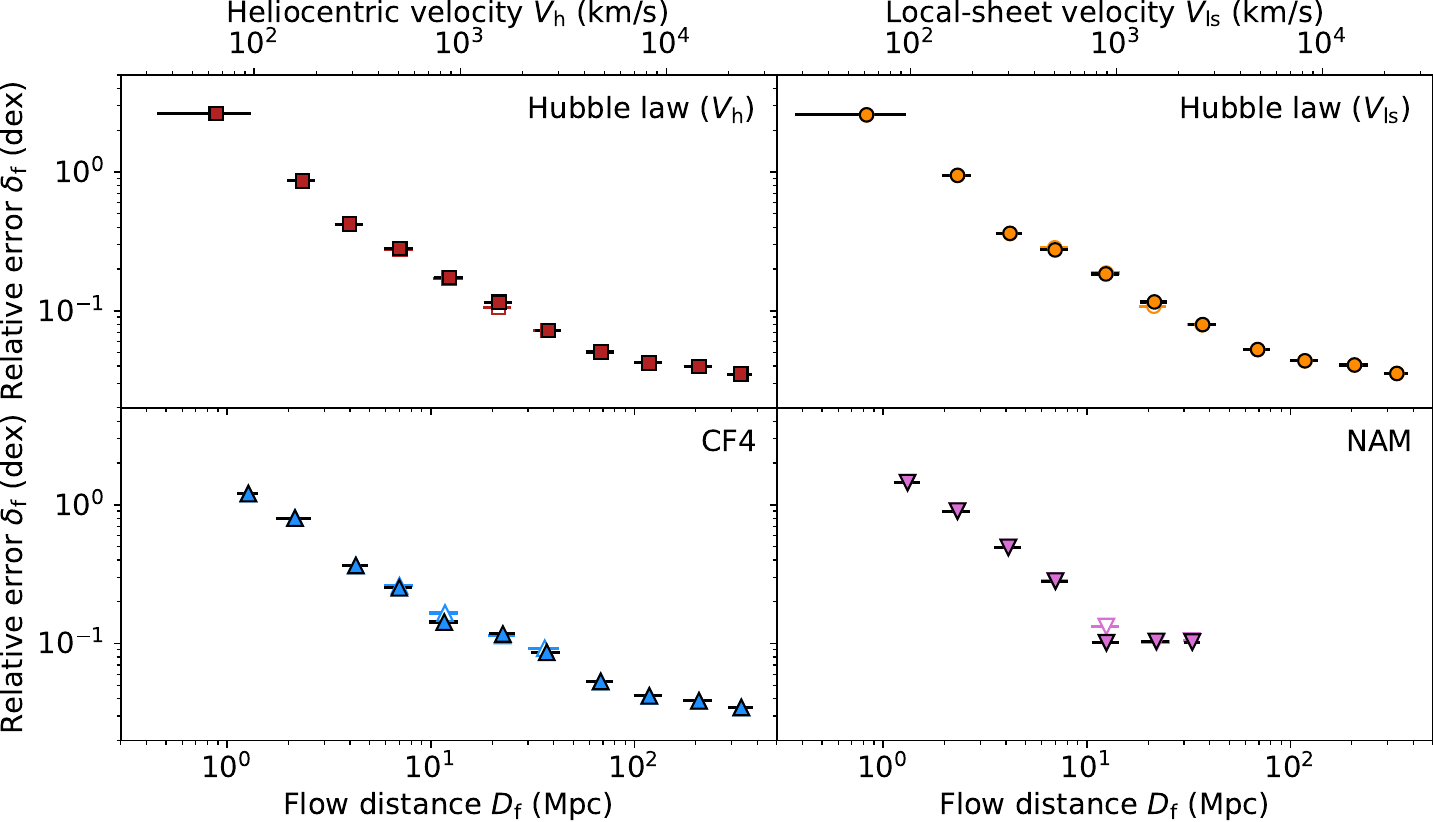}
    \caption{Relative error versus flow distance excluding (filled symbols) and including (open symbols) galaxies within the Fornax ZoI for the four flow methodologies. For the two implementations of the Hubble law, the upper axes show the systemic velocities corresponding to the distances.}
    \label{FigFornaxRelativeErrors}
\end{figure*}

The filled symbols in Fig.~\ref{FigFornaxRelativeErrors} show the relative errors $\delta_\mathrm{f}$ against the flow distances after removing all galaxies in the Fornax ZoI. For comparison, the open symbols show the relative errors without removing the Fornax ZoI. The bins are the same as for Fig.~\ref{FigSDExcVirgo}. The difference between the filled and the open symbols is not noticeable at all for most of the bins. The only noticeable effects are a small increase in relative error in the bin around \SI{20}{\mega\parsec} for the two implementations of the Hubble law and a small decrease in relative error for the bin around \SI{10}{\mega\parsec} for CF4 and NAM. This is in stark contrast to the exclusion of the Virgo ZoI, which decreases the relative error in almost all bins below \SI{30}{\mega\parsec}. Consequently, a separate treatment of uncertainties in the Fornax ZoI is not necessary and Eq.~\ref{EqDBPL} can be used in this region too.

\end{appendix}

\end{document}